\def\year{2021}\relax
\documentclass[letterpaper]{article} 
\usepackage{aaai21}  
\usepackage{times}  
\usepackage{helvet} 
\usepackage{courier}  
\usepackage[hyphens]{url}  
\usepackage{graphicx} 
\usepackage{csquotes}
\usepackage{amssymb}
\usepackage{rotating}
\usepackage{amsmath}
\usepackage{natbib}  
\usepackage{caption} 
\usepackage{array}
\usepackage{flushend}
\usepackage{setspace} 
\usepackage{makecell}
\usepackage[flushleft]{threeparttable}
\usepackage{booktabs,caption}
\usepackage[toc,page]{appendix}
\usepackage{subcaption}
\usepackage{rotating}
\usepackage{multirow}
\usepackage[T1]{fontenc}
\usepackage{xurl}
\usepackage[usenames,dvipsnames]{color}
\usepackage{tikz}

\title{Human Behavior in the Time of COVID-19: Learning from Big Data}
\author {
    Hanjia Lyu, Arsal Imtiaz, Yufei Zhao, Jiebo Luo\\
}
\affiliations {
    University of Rochester\\
    \{hlyu5, aimtiaz2\}@ur.rochester.edu, yzhao87@u.rochester.edu, jluo@cs.rochester.edu\\
}

\begin{document}

\maketitle

\begin{abstract}
Since the World Health Organization (WHO) characterized COVID-19 as a pandemic in March 2020, there have been over 600 million confirmed cases of COVID-19 and more than six million deaths as of October 2022. The relationship between the COVID-19 pandemic and human behavior is complicated. On one hand, human behavior is found to shape the spread of the disease. On the other hand, the pandemic has impacted and even changed human behavior in almost every aspect. To provide a holistic understanding of the complex interplay between human behavior and the COVID-19 pandemic, researchers have been employing big data techniques such as natural language processing, computer vision, audio signal processing, frequent pattern mining, and machine learning. In this study, we present an overview of the existing studies on using big data techniques to study human behavior in the time of the COVID-19 pandemic. In particular, we categorize these studies into three groups - using big data to \textit{measure}, \textit{model}, and \textit{leverage} human behavior, respectively. The related tasks, data, and methods are summarized accordingly. To provide more insights into how to fight the COVID-19 pandemic and future global catastrophes, we further discuss challenges and potential opportunities.
\end{abstract}

As of October 2022, the COVID-19 pandemic has caused more than 600 million confirmed cases and more than six million deaths all over the world,\footnote{\url{https://covid19.who.int/}} bringing unprecedented damage and changes to the human society~\citep{singh2020covid,verma2020impact, maital2020global}. Meanwhile, human behavior such as mobility, and non-pharmaceutical interventions (NPI) also affect the development of the pandemic~\citep{chang2021mobility, hirata2022did}. Therefore, to fight against the COVID-19 pandemic, not only knowing the virus itself is important, it is also critical to better understand the interplay between human behavior and COVID-19. 

Despite the damage and negative impact, the COVID-19 pandemic also indirectly accelerates the adoption of digital tools across various fields globally such as medicine, clinical research, epidemiology, education, and management science~\citep{gabryelczyk2020has, xu2021artificial}, which provides a distinctive opportunity to study human behavior through the lens of big data techniques. 

Several prior surveys have been conducted on the topic of using big data applications to combat the COVID-19 pandemic. \citet{bragazzi2020big} discussed the potential applications of artificial intelligence and big data techniques of different time scales from short-term such as outbreak identification to long-term such as smart city design to manage the pandemic. \citet{haleem2020significant} summarized eight applications of big data in the COVID-19 pandemic. \citet{alsunaidi2021applications} highlighted several domains that manage and control the pandemic on the basis of big data analysis. However, they did not delve into the discussion about using big data techniques to learn \textbf{human behavior} in the context of COVID-19.

There have been some surveys that investigate both big data and human behavior. They mostly focus on \textbf{specific topics} such as social media and misinformation~\citep{greenspan2021pandemics, huang2022social}, technologies and their effects on humans~\citep{vargo2021digital, agbehadji2020review, haafza2021big}, social and behavioral science~\citep{xu2021artificial, sheng2021covid}. These surveys have presented an important overview of each \textbf{independent} research topic within the human behavior domain. However, it is also critical to understand big data and human behavior holistically. Therefore, we aim to present a review of how big data technologies can be used to learn human behavior in the context of the COVID-19 pandemic \textbf{in a larger scope} compared to previous surveys.

We use Google Scholar, PubMed, and Scopus to search for related papers that were published between 2020 and 2022. Keywords include ``COVID-19'', ``human behavior'', ``big data'', ``deep learning'', ``machine learning'', and ``artificial intelligence''. Surveys, comments, and studies that do not use any specific big data techniques are excluded. After reviewing the related tasks, data, and methods of these papers, we categorize them into three groups based on how these big data techniques are used to study human behavior. Specifically, we include studies that use big data techniques to \textit{measure}, \textit{model}, and \textit{leverage} human behavior.

\begin{figure*}[t]
    \centering
    \includegraphics[width = \linewidth]{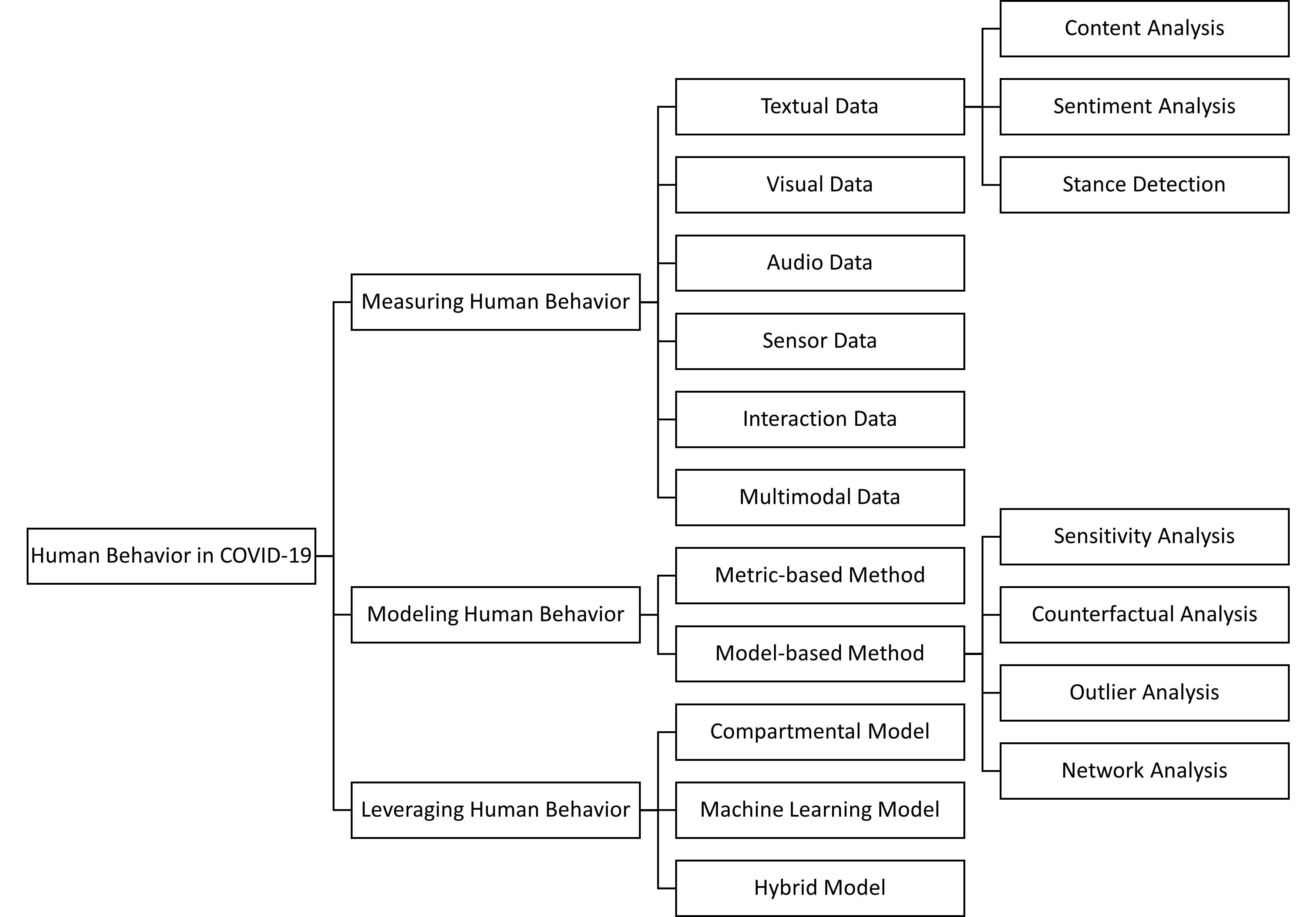}
    \caption{The overall categorization of the reviewed studies. }
    \label{fig:cat}
\end{figure*}

The remainder of the survey is organized as follows: In Section 2, we discuss how big data techniques can be used to \textit{measure} human behavior with respect to various \textit{data types} (e.g., textual, visual, audio). In Section 3, we illustrate two types of methods including the metric-based and the model-based that can be used to \textit{model} human behavior. In Section 4, we introduce three lines of models including compartmental models, machine learning models, and hybrid models that have been heavily used to \textit{leverage} human behavior in the context of COVID-19. For each category (i.e., measure/model/leverage), we summarize the related tasks, data, and methods accordingly (Figure~\ref{fig:cat}), and provide detailed discussions about related challenges and potential opportunities. Finally, we discuss other challenges and open questions in Section 5, and conclude our study in Section 6.

\section{Using Big Data to Measure Human Behavior in COVID-19}
Unlike conventional methods that measure human behavior via questionnaires, with the increasing usage of digital technologies, measuring human behavior from raw data such as video and text has been widely used during the COVID-19 pandemic. Using big data to \textit{measure} human behavior intends to \textit{mine interesting patterns from raw data that records human behavior with big data techniques}. The goal is to map the raw data which is often unstructured to measurements that quantitatively describe human behavior. We primarily, but not exclusively, discuss six groups of studies based on the data types including textual, visual, audio, sensor, interaction, and multimodal.

\subsection{Textual Data}
For textual data, researchers focus on the content that is generated when people use the Internet to measure human behavior. It can be the queries people use to search for health-related information, and the posts people write on social platforms. During the pandemic, people have been found to increase their social media use for multiple reasons such as seeking social support~\citep{rosen2022social}, expressing opinions~\citep{lyu2022social,tahir2022improving, wu2021characterizing}, and obtaining health information~\citep{tsao2021social,rabiolo2021forecasting}. The main format of social media posts is textual data. The growing amount of online textual data presents a channel to passively observe human behavior more efficiently at a large scale. In this section, we discuss three analysis frameworks that can be used to measure human behavior from textual data including content analysis, sentiment analysis, and stance detection. Note that although we focus on the studies that use these three frameworks on textual data since textual data is predominant, these frameworks can also be applied to other data types such as visual data. 

\subsubsection{Content Analysis}
Content analysis includes ``any technique for making inferences by objectively and systematically identifying specified characteristics of message''~\citep{holsti1969content}. During the pandemic, it is primarily applied to online discourse to capture topics and measure people's opinions toward the disease, public policies, and societal changes. 

For instance, \citet{evers2021covid} extracted and compared the word frequencies in four Internet forums 70 days before and 70 days after the date Former US President Trump declared a national emergency. Word frequencies were used to measure search interest. Words that are relevant to the Theory of Social Change, Cultural Evolution, and Human Development, and have a relatively narrow range of meanings were included. Across four forums, they found that words or phrases that are related to subjective mortality salience (e.g., ``cemetery'', ``survive'', ``death''), engagement in subsistence activities (e.g., ``farm'', ``garden'', ``cook''), and collectivism showed increases (e.g., ``sacrifice'', ``share'', ``help''), suggesting that human may shift their behavior according to the level of death and availability of resources. \citet{cao2021analysis} leveraged Latent Dirichlet Allocation (LDA)~\citep{blei2003latent} to mine the various topics in the Sina Weibo posts that show attitudes toward the lockdown policy in Wuhan. LDA is a generative probabilistic model of a corpus. \citet{cao2021analysis} first used a Chinese word segmentation tool to segment sentences and removed stopwords. Next, they calculated the Term Frequency-Inverse Document Frequency (TF-IDF). Finally, an LDA model was applied and trained. The optimal number of topics was decided based on the perplexity of each topic. Eight topics were identified including (1) daily life under lockdown, (2) medical assistance, (3) traffic and travel restrictions, (4) epidemic prevention, (5) material supply security, (6) praying for safety, (7) unblock of Wuhan, and (8) quarantine and treatment. They found that people's emotions changed due to different topics. They argued that with the development of the pandemic, emotions involving more uncertainty such as hope declined while emotions involving more certainty such as admiration and joy increased.

\subsubsection{Sentiment Analysis}
Sentiment analysis is to detect sentiment expressed in the content toward an entity~\citep{medhat2014sentiment}. Textual sentiment analysis is a well-studied classification task that intends to estimate whether the sentiment expressed in text is positive or negative. It has been leveraged in multiple applications including measuring the polarity of people's opinion~\citep{liu2021clothing, sanders2021lessons, yeung2020face,chen2021fine} and quantifying mental well-being~\citep{zhang2021monitoring, aslam2020sentiments}. There are two major ways to estimate the sentiment from text. One is lexicon-based, and the other is machine learning-based. Lexicon-based methods require a pre-labeled dictionary that maps each word to a polarity score indicating positive or negative, or other emotions. \citet{barnes2021understanding} used LIWC (Linguistic Inquiry and Word Count)~\citep{pennebaker2001linguistic} which captures the psychological states and sentiment from the text to detect terror states. In particular, they focused on anxiety, death, religion, reward, affiliation, and social processes. They compared the number of new cases and the number of tweets in which anxiety and death-related words were detected. They found that they evolved consistently. The other is machine learning-based methods which intend to learn a function to predict the sentiment of a piece of text. One important difference is that machine learning-based methods are more robust to estimate sentiment in different contexts~\citep{jaidka2020estimating}. \citet{choudrie2021applying} fine-tuned a pre-trained language model and applied it to over two million tweets from February to June 2020 in the context of COVID-19 to assess public sentiment toward various government management policies. They complemented the existing emotions that were archived by the research conducted before the pandemic~\citep{jain2017extraction} with the emotions that are more pandemic-specific including anger, depression, enthusiasm, hate, relief, sadness, surprise, and worry. The accuracy and average Matthew correlation coefficient (MCC)~\citep{chicco2020advantages} of the sentiment classification are 0.80 and 0.78, respectively. 


\subsubsection{Stance Detection}
Although sentiment analysis measures the sentiment toward an entity, it does not necessarily capture whether a person supports or disagrees with any opinions. A piece of text containing multiple positive sentiment-related words might be classified as positive by sentiment analysis tools. However, the person who posts it can still be opposed to the text's view. Thus, stance analysis or stance detection is more suitable to tackle this problem. Stance detection is to classify the stance of the author~\citep{kuccuk2020stance}. One of the many applications of stance detection for COVID-19 is detecting public response to policies (e.g., NPIs) or major events. For instance, \citet{tahir2022improving} combined online (i.e., tweets and user profile descriptions) and offline data (i.e., user demographics) to measure people's vaccine stance. They designed a multi-layer perceptron with six layers that incorporates both the semantic features from tweets and user profile descriptions and one-hot representations of user demographics including state, race, and gender. The model was trained and evaluated using a dataset of 630,009 tweets with 69,028 anti-vaccine and 560,981 pro-vaccine. The opinion (i.e., pro-vaccine or anti-vaccine) was determined by hashtags including \textit{\#NoVaccine}, \textit{\#VaccinesKill}, \textit{\#GetVaccinated}, and \textit{\#VaccinesWork}.  \citet{lyu2022social} adopted a human-guided machine learning framework to classify tweets into pro-vaccine, anti-vaccine, and vaccine-hesitant in an active learning manner. Out of the 2,000 initially labeled tweets, only 16 percent are relevant to opinions. To accelerate the annotation and model construction process, they then labeled tweets iteratively. In each iteration, the model was trained with all labeled data. Next, it ranked the rest of the unlabeled data and output tweets that were most likely relevant to opinions. Human annotators manually labeled these tweets and merged this newly labeled batch to the labeled corpus. In this way, the model and human worked together to actively search for relevant tweets, increasing the efficiency of the modeling process. They applied the trained model to the tweets of 10,945 unique Twitter users and found that opinion on the COVID-19 vaccine varies across people of different characteristics. More importantly, a more polarized opinion is observed among the socioeconomically disadvantaged group. 

Stance detection is complex because it often requires extracting language patterns that are informative in terms of specific topics. However, labeling a brand new dataset from scratch is still time-consuming and expensive even with active learning. \citet{miao2020twitter} showed that a small amount of data for the task of interest can significantly improve the performance of models that are trained with existing datasets for stance detection. Therefore, to facilitate stance detection in the context of COVID-19, \citet{glandt2021stance} annotated a COVID-19 stance detection dataset - COVID-19-Stance, that is composed of 6,133 tweets expressing people's stance toward ``Anthony S. Fauci, M.D.''. ``Keeping Schools Closed'', ``Stay at Home Orders'', and ``Wearing a Face Mask''.

\subsection{Visual Data}

While computer vision techniques have been primarily adapted in healthcare, methods, especially object detection, have been applied to measuring human behavior~\citep{cvhealthcare1, cvhealthcare2}. We discuss several object detection applications in the context of COVID-19 based on the scales of the objects. 
 
 We first discuss the applications of computer vision that consider faces from a near distance. To help conduct prevention advised by the health institutions during the COVID-19 pandemic, computer vision systems are designed to detect whether people properly wear masks, avoid touching their faces, and keep away from crowds. The goal of a face mask detection system is two-fold - detecting whether the person is wearing a face mask or not, and detecting if they are wearing the mask properly. \citet{eyiokur2022unconstrained} collected and annotated a face dataset from the web, namely \textbf{I}nteractive \textbf{S}ystems \textbf{L}abs \textbf{U}nconstrained \textbf{F}ace \textbf{M}ask \textbf{D}ataset (ISL-UFMD). ISL-UFMD contains 21,816 images with a great number of variations. Compared to previously proposed mask detection datasets~\citep{cabani2021maskedface,yang2021vision}, there is a larger variety of subjects' ethnicity, and image conditions such as environment, resolution, and head pose in ISL-UFMD. \citet{eyiokur2022unconstrained} achieved an accuracy of 98.20\% with Inception-v3~\citep{Inception}. They also implemented an Efficient-Net~\citep{efficientnet} model to detect face-hand interaction on the \textbf{I}nteractive \textbf{S}ystems \textbf{L}abs \textbf{U}nconstrained \textbf{F}ace \textbf{H}and \textbf{I}nteraction \textbf{D}ataset (ISL-UFHD) and achieved an accuracy of 93.35\%. Unlike the existing face-hand interaction dataset~\citep{beyan2020analysis} which is limited with respect to the number of subjects and the controlled data collection conditions, ISL-UFHD was collected from unconstrained real world scenes.

COVID-19 has also impacted the education system as in-person learning was transited into online learning to slow down the spread of the pandemic~\citep{basilaia2020transition}. \citet{bower2019technology} stated that there might be negative learning outcomes if students do not feel a sense of cognitive engagement through online learning. To tackle this problem, \citet{bhardwaj2021application} proposed a deep learning framework that measures students' emotions in real-time such as anger, disgust, fear, happiness, sadness, and surprise. The ``FER-2013'' facial image dataset~\citep{goodfellow2013challenges} (consisting of 35,887 $48\times48$ gray scaled images) and the Mean Engagement Score (MES) dataset were used. The MES dataset was collected and curated through a survey of 1,000 students over one week. A combination of the Haar-Cascade Classifier~\citep{viola2001rapid} and a CNN classifier was used for engagement detection. The proposed method achieved an accuracy of 93.6\% on engagement detection. Combined with emotion recognition, this system helps teachers assess students' level of engagement during online classes.

Secondly, visual data has been recorded in urban settings - office spaces, campuses, busy streets, and so on to study the movement of people and whether they adhere to COVID-19 guidelines. \citet{saponara2021implementing} presented a real-time deep learning system that measures distances between people using state-of-the-art YOLO~\citep{redmon2016you} object detection model. The system is mainly composed of three steps. First, an object detector is applied to thermal images or video streams for people detection. The number of people is counted. Second, they compute the distance between the centroid of the bounding boxes containing people. Finally, adherence to social distancing guidelines is assessed based on the number of detected people and the distances. The model attained an accuracy of 95.6\% on a dataset that consists of 775 thermal images of humans in various scenarios, and 94.5\% on Teledyne FLIR dataset\footnote{\url{https://www.flir.com/oem/adas/adas-dataset-form/}} that consists of 800 IR images. More importantly, experimental results further showed that their proposed system handled more frames per second than R-CNN~\citep{girshick2014rich} and Fast R-CNN~\citep{ren2015faster}, suggesting its applicability in real-time detection. \citet{shorfuzzaman2021towards} approached the problem with the vision of it being adaptable to surveillance cameras set up in smart cities. Their proposed methodology involves a perspective transformation of the image to a bird's eye view prior to feeding it to the neural network. They adopted and evaluated object detection models - Faster R-CNN~\citep{ren2015faster}, YOLO~\citep{redmon2016you}, and SSD~\citep{liu2016ssd} on the publicly available Oxford Town Center dataset~\citep{benfold2011stable} which contains videos of urban street setting with a resolution of $1920 \times 1080$ sampled at 25 FPS. YOLO performs the best amongst the three models with an mAP and IoU of 0.847 and  0.906, respectively. The current implementation detects pedestrians in a region of interest using a fixed monocular camera and estimates the distance in real-time \textbf{without} recording data.

Last but not the least, satellite imagery has applications of providing overarching perspectives of activities from a larger scale.  \citet{satellite} proposed a methodology to study human and economic activity and support decision-making during the pandemic by extracting temporal information from satellite imagery using a lightweight ensemble of CNNs. The system trained on the xView~\citep{xview} and fMoW~\citep{fMOW} datasets achieved an mAP of 0.94 on identifying aeroplanes and 0.66 on identifying smaller vehicles. A sequence of images allows the model to keep track of these vehicles and extract temporal information about the location. The study provides observation of human activity during the stay-at-home order and economic activity carried out in North Korea, Russia, Germany, etc. \citet{satellite} also compared images before and after the pandemic. For instance, their system automatically detected a decrease in the number of planes in activity at the Salt Lake City International Airport before and after the outbreak. The images captured in other places of interest such as supermarkets and toll booths indicated compliance with the stay-at-home orders.

\subsection{Audio Data}
The research community has been using speech and audio data to combat the COVID-19 pandemic through automatic recognition of COVID-19 and its symptoms such as breathing, coughing, and sneezing~\citep{schuller2021covid}. In our survey, we focus on the existing studies that use audio data to measure human behavior instead of diagnosis.

Videos and speeches are explored and analyzed to understand the sentiment and mental conditions during the pandemic. Social vloggers posted videos on social platforms to share their opinions and experiences during the quarantine. \citet{feng2020exploring} collected 4,265 YouTube videos to track vloggers' emotions and their association with pandemic-related events. There are two major issues regarding the collected audio data. First, they may contain multiple speakers. Second, the non-speech segments are mostly noise and music background. To address these two issues, they manually labeled a 5-second reference audio of the target speaker of each video. Speaker diarization was then conducted by calculating the similarity between the reference audio and each target analysis window. The similarity metric was computed by a speaker verification model~\citep{wan2018generalized}. The threshold for the similarity score was set to 0.65 and the window size was set to 125 milliseconds. Linear regression was leveraged to interpret emotions based on four prosodic features including loudness, zero-crossing rate, jitter, and shimmer. Stress levels and emotions were also measured via regression analysis based on speech samples collected over phone call conversations~\citep{konig2021measuring}. Participants were asked to describe something emotionally neutral, negative, and positive. Each description should last about one minute. Complicated machine learning tools were used as well. \citet{han2020early} applied support vector machines (SVMs) to the acoustic features extracted from the speech recordings of 52 COVID-19-diagnosed patients from two hospitals in Wuhan, China to measure the anxiety level. \citet{elsayed2022speech} proposed a hybrid model of the gated recurrent neural networks (GRU) to evaluate emotional speech sets with an accuracy of 94.29\%. 

Apart from sentiment analysis and stress level assessment, audio data are also applied in the measures of mechanics of phonation and speech intelligibility. \citet{deng2022modeling} investigated the effects of mask-wearing and social distancing on phonation and vocal health. While simulations were performed for vowel phonemes with different masks or multiple mask layers worn, a three-mass body-cover model of the vocal folds (VFs) coupled with the subglottal and supraglottal acoustic tracts was modified to incorporate mask and distance-dependent acoustic pressure models. The acoustic wave propagation was modeled with the wave reflection analog (WRA) method. Wearing masks might reduce intelligibility. However, their study provides several practical insights into how this effect can be mitigated while allowing masks to keep their role in preventing airborne disease transmission. They found that light masks are preferable to heavy masks for the same particle filtration properties and the negative effects of wearing masks on intelligibility can be greatly reduced in a low-noise environment.

Similarly, \citet{knowles2022impact} measured speech intensity, spectral moments, and spectral tilt and energy of sentences read by 17 healthy talkers with three types of mask-wearing by linear mixed effects regression. With statistical analysis on speech-related and phonation-related variables, \citet{porschmann2020impact} analyzed the impact of face masks on voice radiation with audio data generated by a mouth simulator covered with six masks. \citet{bandaru2020effects} assessed the speech perception based on the results of 20 participants' speech audiometry testing. 

There are other aspects of human behaviors being investigated with the audio data as well. For instance, \citet{MOHAMED2022108361} examined whether voice biometrics can be leveraged in mask detection, that is, classifying whether or not a speaker is wearing a mask given a piece of audio data. They reviewed the face mask detection approaches via voice in the Mask Sub-Challenge (MSC) of the INTERSPEECH 2020 COMputational PARalinguistics challengE (ComParE). There are mainly two groups of methods:(1) models using phonetic-based audio features, and (2) frameworks that combine spectrogram representations of audio and CNNs. They also found that fusing the approaches can improve classification performance. Finally, they implemented an Android-based smartphone app based on the proposed model for mask detection in real time.

\subsection{Sensor Data}
With the development of the Wireless Sensor Network, sensing devices that measure physical indicators have been applied to multiple areas such as ambient intelligence~\citep{haque2020illuminating} and environmental modeling~\citep{wong2016real}. Because of the reliability, accuracy, flexibility, cost-effectiveness, and ease of deployment of sensor networks~\citep{10.1145/565702.565708}, researchers mine sensor data to measure human behavior in the context of COVID-19. Body temperature monitoring has been implemented as one of the prevention strategies during the pandemic. However, the accuracy of temperature measurement is not consistent in some cases. For instance, the surrounding environment such as ambient temperature and humidity may easily impact the data. Moreover, body motion influences temperature~\citep{zhang2021body}. To alleviate these issues, \citet{zhang2021body} recruited 10 participants to wear a wearable sensor that contains a three-axis accelerometer, a three-axis gyroscope, and a temperature sensor. The three-axis acceleration, three-axis angular velocity, and body surface temperature were recorded at 50 Hz while these participants were sitting, walking, walking upstairs, and walking downstairs. Using the seven-dimensional sensor data (i.e., acceleration, velocity, and temperature) as input, \citet{zhang2021body} intended to employ conventional machine learning, deep learning algorithms, and ensemble methods to predict human activities. In particular, they applied conventional machine learning algorithms such as support vector machine, logistic regression, and so on. Stacked denoising autoencoder~\citep{vincent2010stacked} was leveraged. Ensemble methods that were used include random forest~\citep{breiman2001random}, extra trees~\citep{geurts2006extremely}, and deep forest~\citep{zhou2017deep}. Only KNN and the ensemble methods performed better in the human activity recognition task. The experimental results indicated that body temperature is lower when a human is involved in dynamic activities (e.g., walking) compared with static activities (e.g., sitting), which may potentially cause misses in fever screening for COVID-19 prevention. \citet{sardar2022mobile} extended human activity recognition to handwashing, standing, hand sanitizing, nose-eyes touching, handshake, and drinking. 

Except for wearable devices, researchers have also used mobile phone sensing to measure human behavior. In \citet{nepal2022covid}, 220 college students were recruited and asked to keep a continuous mobile sensing app installed and running. \citet{nepal2022covid} tried to understand the objective behavioral changes of students under the influence of the COVID-19 pandemic. The app records features including phone usage, mobility, physical activity, sleep, semantic locations, audio plays, and regularity. Students were found to travel significantly less during the first COVID-19 year but use their phones more. Fully Convolutional Neural Networks~\citep{wang2017time} was applied to predict students' COVID-19 concerns solely based on the sensor data. The model achieved an AUROC of 0.70 and an F1 score of 0.71. The number of unique locations visited, duration spent at home, running duration, audio play duration, and sleep duration were the most important features when predicting students' concerns regarding COVID-19.

Despite the merits discussed before, issues such as missing data, outliers, bias, and drift can occur in sensor data collection~\citep{teh2020sensor}. The errors at the individual level may be small, however, when aggregated, they can be enlarged and reduce the performance of large-scale predictive models~\citep{ienca2020responsible}. In addition to enhancing the data collection process, error detection and correction pipelines~\citep{teh2020sensor} can be applied before using the sensor data to measure human behavior.

\subsection{Interaction Data}
Interaction data such as the number of logins, the number of posts, etc. records how people use and interact with the digital systems. Researchers apply data mining techniques to this type of data to recognize interesting patterns and measure human behavior. One of the major applications during the COVID-19 pandemic is evaluating the quality of online learning~\citep{dascalu2021before, dias2020deeplms}. Video conferencing platforms such as Zoom and MS Teams are heavily used for online learning. The quality of engagement is important to the quality of online learning~\citep{herrington2007immersive}. To provide instructors and learners with supportive tools that can improve the quality of engagement, \citet{dias2020deeplms} applied a deep learning framework that leverages users' history interaction data in a learning management system to provide real-time feedback. The learning management system records 110 types of interaction data which are processed by a fuzzy logic-based model~\citep{dias2013fuzzyqoi} to measure the quality of interaction (QoI) at each timestamp. Next, an LSTM model is trained with this time series data. The difference between the predicted QoI at timestamp $k+1$ and the real QoI at timestamp $k$ is used as the metacognitive stimulus to improve the quality of learning. A positive difference is considered rewarding feedback while a negative difference warning feedback. They applied the proposed method to one database before and two databases during the COVID-19 pandemic. The average correlation coefficient between ground truth and predicted QoI values is no less than 0.97 ($p<0.05$).

\subsection{Multimodal Data}
Multimodal human behavior learning uses at least two types of aforementioned data. The motivation is straightforward, that is, to exploit complementary information of multiple modalities to improve robustness and completeness~\citep{baltruvsaitis2018multimodal}. Social platforms have abundant multimodal data. The publicly available profile image uploaded by a user may contain the demographic information of this user~\citep{xiong2021social,zhang2021understanding,lyu2020sense}. This side information can potentially improve the performance of measuring human behavior. For instance, \citet{zhang2021monitoring} fine-tuned an XLNet model~\citep{yang2019xlnet} and applied it to social media posts to predict whether or not the author of the posts has depression. To improve the detection performance, they obtained user demographics by leveraging a multimodal analysis model - M3 inference model~\citep{wang2019demographic} which predicts the age and gender of the user based on the user's profile image, username, and description. The fusion model that uses all features including the depression score estimated by the XLNet model and the demographic inferred by the M3 inference model outperforms other baselines and achieves an accuracy of 0.789 on the depression detection task. They further demonstrated the ability of their model in monitoring both group- and population-level depression trends during the COVID-19 pandemic. All tweets posted from January 1 to May 22, 2020 of 500 users were used to measure users' depression levels. They found that the depression levels of the users in the depression groups were substantially higher than that in the non-depression group. The depression levels of both groups roughly corresponded to three major events in the real world including (1) the first case of COVID-19 confirmed in the US (January 21), (2) the US National Emergency announcement (March 13), and (3) the last stay-at-home order issued (South Carolina, April 7).

\subsection{Challenges and Opportunities}
Although big data techniques have shown promising performance in measuring human behavior in the applications during the COVID-19 pandemic, there are still several challenges and room for improvement in terms of the \textbf{three characteristics of the spread of COVID-19} including (1) its unprecedented global influence, (2) disparate impacts over people across various race, gender, disability, and socioeconomic status~\citep{van2020covid}, and (3) rapidly evolving situations.
\begin{itemize}
    \item COVID-19 has impacted the society and economy worldwide. The scale and scope of the impact of COVID-19 have led to an unprecedented increase in the amount of online health discourse~\citep{wu2021characterizing,evers2021covid}. The data distributions may be \textit{entirely different} for research questions on a similar topic before and during the pandemic, thus requiring certain adaptations. For instance, both regarding using Twitter data to understand vaccine hesitancy, the majority of the collected tweets in \citet{tomeny2017geographic} are relevant to vaccine opinions while the majority of the collected tweets in \citet{lyu2022social} are irrelevant. To infer opinions from tweets, they both labeled a small batch of data first, then trained a machine learning model. Given the low percentage of relevant data in \citet{lyu2022social}, it would have been inefficient and led to poor performance of opinion inference if the same data process pipeline was applied.

    Distribution shift may also compromise the \textit{robustness of off-the-shelf methods} in measuring human behavior in the era of COVID-19. Specifically, many studies use off-the-shelf sentiment analysis tools such as VADER (Valence Aware Dictionary and sEntiment Reasoner)~\citep{hutto2014vader} and LIWC~\citep{pennebaker2001linguistic} to estimate human sentiment and emotion based on lexicon scores that were \textit{labeled well before the pandemic}. These methods may not yield robust sentiment estimates in the context of COVID-19 which has greatly impacted human society, as evidence has shown that the robustness of lexicon-based sentiment analysis tools can be worsened compared to machine learning-based methods because of regional cultural and socioeconomic differences in language use~\citep{jaidka2020estimating}.

    The change in how data is generated in the COVID-19 context also calls for model adaptation. For instance, \citet{de2011facial} found that facial expressions are important in distinguishing different types of sentences in signed languages. The performance of models in sign language recognition may be undermined because of the mask order where the movement of jaws and cheeks are covered. Future work may need to adapt the models to maintain the performance. 
    
    \item The uneven influences of the COVID-19 pandemic over people across various characteristics~\citep{van2020covid} strongly suggests that understanding and investigating the disparities among different populations~\citep{lyu2022social} is essential. This phenomenon also contributes to algorithmic bias~\citep{walsh2020stigma}. For instance, \citet{aguirre-etal-2021-gender} analyzed the fairness of depression classifiers trained on Twitter data in terms of different gender and racial demographic groups. They found that models perform worse for underrepresented groups. How to pursue fairness in measuring human behavior with respect to different populations in the context of COVID-19 remains an open question.
    
    \item For visual data, systems that leverage state-of-the-art YOLO algorithms~\citep{redmon2016you} perform poorly on small targets, thus limiting the application to a particular scale of human faces in images~\citep{yolochallenge1}. Furthermore, \citet{maskdetectorchallenge2} stated that the time performance of CNN models decreases with an increase in model parameters without GPU support. There is a trade-off between time performance and accuracy, which is critical to the deployment of such applications in real-time systems. Another reason that real-time systems are important is data privacy issues. For instance, individuals may not accept being monitored by surveillance systems. To avoid violating privacy when applying models to the video surveillance data, the system should be fast in inference in order not to record or store data~\citep{shorfuzzaman2021towards}. Moreover, designing models using techniques such as knowledge distillation~\citep{gou2021knowledge} may be helpful for developing real-time systems. Knowledge distillation compresses information from a large teacher model and learns a small student model. The teacher model is often trained with data at a large scale and is hard to deploy. In contrast, the student model is learned via knowledge distillation and is lightweight. 

    Additionally, there exists another avenue for future research regarding real-time systems in the context of COVID-19. Based on the characteristics of different types of data and the cost of implementation, we should be more flexible in choosing the optimal data to address problems in fighting against the COVID-19 pandemic. As has been shown, computer vision techniques can be used for monitoring social distancing. However, it is often expensive to train these deep learning algorithms. \citet{li2021covid} proposed an alternative way to count people within a given place by using a lightweight supervised learning algorithm to exploit WiFi signals, which is low-cost and non-intrusive. 
    
    
\end{itemize}

\section{Using Big Data to Model Human Behavior in COVID-19}
After obtaining human behavior measurements either from questionnaires, raw data, or both, big data techniques are employed to \textit{model} and \textit{analyze} human behavior. The goal of modeling human behavior is to \textit{characterize human behavior and understand the relevance between human behavior and other variables}. In this section, we discuss existing studies in analyzing human behavior based on the method types including metric-based and model-based.

\subsection{Metric-based Method}
This line of methods focuses on using metrics to characterize human behavior and describe its relevance to other variables. The metrics can be calculated directly based on the data distributions, which does not require model construction. 

Well-studied statistical methods such as correlation analysis~\citep{ezekiel1930methods}, Chi-square test~\citep{mchugh2013chi}, Student's t test~\citep{kim2015t}, and ANOVA (Analysis of Variance)~\citep{st1989analysis} have been widely applied to behavioral science in terms of COVID-19. The measure of the mutual dependence between two random variables is defined as mutual information~\citep{shannon1948mathematical}. It has been used to analyze the relationship between the growth of COVID-19 cases and human behavior such as isolation, wearing masks outside home, and contact with symptomatic persons. \citet{tripathy2021prediction} calculated the mutual information regression score to assess the feature importance and found high relevance between case increase and behavioral features including avoiding gatherings, avoiding guests at home, wearing masks at public transport, and willingness to isolate.

The counts of different frequent patterns have been used to characterize human behavior. Moreover, the confidence, support, and lift of association rules that are generated from the frequent patterns are used to indicate the relevance between human behavior and other variables. \citet{urbanin2022social} applied the Apriori algorithm~\citep{agrawal1994fast} to the survey responses of 10,162 participants to generate association rules with respect to people's social and economic conditions and the reports of protective behaviors including the use of protective masks, distancing by at least one meter when out of the house, and handwashing or use of alcohol. Increased use of video conferencing platforms and the possibility to work from home are both associated with better self-care behavior. Fear of economic struggle is associated with preventive behavior. The generated association rules provide insights into how different stages of the pandemic shape human behavior.

\subsection{Model-based Method}
These approaches emphasize on building models to analyze human behavior and explain the relationship between it and other variables. We discuss four analysis frameworks including sensitivity analysis, counterfactual analysis, outlier analysis, and network analysis. Note that although we focus on the studies that use these four frameworks on model-based methods, some of these frameworks can also be applied to metric-based methods. 

\subsubsection{Sensitivity Analysis}
Sensitivity analysis investigates the relationship between the variation in input values and the variation in output values~\citep{saltelli2002sensitivity}. \citet{christopher2002identification} have systematically reviewed several sensitivity analysis methods. In our study, we focus on discussing its applications in analyzing human behavior in the context of COVID-19. 

\citet{ramchandani2020deepcovidnet} performed sensitivity analysis~\citep{samek2018explainable} to explain the feature importance. The assumption is that the model is most sensitive to the value variation of the most important features. More specifically, they first evaluate the accuracy of the model on a small subset of the training data. Next, for each feature, they randomized its value and evaluated the accuracy of the model on the same subset again. The importance of the feature is assigned based on how much the accuracy decreases due to the randomization. The lower the accuracy is, the more important the feature is. They found that incoming county-level visits are one of the most important features in predicting the growth of COVID-19 cases.

\citet{Rodríguez_Tabassum_Cui_Xie_Ho_Agarwal_Adhikari_Prakash_2021} proposed a variation of the sensitivity analysis where they removed the features instead of randomizing their values. Moreover, they conducted a two-sample t-test with the null hypothesis that the performance does not significantly change after they drop the features. Instead of simply using accuracy as the indicator, \citet{bhouri2021covid} employed a revised sensitivity measure~\citep{campolongo2007effective} which was originally proposed by \citet{morris1991factorial} to quantify the feature importance. This measure is calculated based on the distribution of the elementary effect of input features.

Other statistical models such as multinomial logistic regression~\citep{kwak2002multinomial}, structural equation modeling (SEM)~\citep{hox1998introduction}, and negative binomial regression~\citep{hilbe2011negative} are also applied to analyze the relationship between human behavior and other variables. 

 \citet{yang2020consumption} applied SEM to the survey responses of 512 participants to explore the relationship between COVID-19 involvement and consumer preference for utilitarian versus hedonic products. They found that COVID-19 involvement is positively related to consumer preference for utilitarian products and attribute it to the mediated effects of awe, problem-focused coping, and social norm compliance. \citet{lyu2022misinformation} conducted the Fama-MacBeth regression with the Newey-West adjustment on nearly four million geotagged English tweets and the COVID-19 data (e.g., vaccination, confirmed cases, deaths) to analyze how people react to misinformation and fact-based news on social platforms regarding the COVID-19 vaccine uptake, and discovered that the negative association between fact-based news and vaccination intent might be due to a combination of a larger user-level influence and the negative impact of online endorsement.



\subsubsection{Counterfactual Analysis}
Counterfactual analysis focuses on simulating the target variable with hypothetical input data. It intends to investigate the effects of changes in certain variables on the output variable, which is similar to sensitivity analysis. However, sensitivity analysis pays more attention to the input side while counterfactual analysis focuses on the output side. Given this characteristic, counterfactual analysis has been widely applied to policy design as it simulates the effect of policy interventions~\citep{lyu2022social, chang2021mobility, rashed2021one, hirata2022did}. For instance, \citet{hirata2022did} conducted the counterfactual analysis to investigate the effects of people's activity during holidays and vaccination on the COVID-19 spread. In particular, they compared three time series. The first time series is the estimated new daily positive cases after July 23, 2021, assuming vaccination was not conducted. The second time series is the estimated new daily positive cases after July 23, 2021, assuming the mobility was identical to the mobility before the pandemic. The third time series is the observed new daily positive cases after July 23, 2021. By comparing them, they found that both human mobility and vaccination have an impact on the spread of COVID-19. The effect of vaccination is larger than the effect of human mobility.

\subsubsection{Outlier Analysis}
The data objects that do not comply with general behavior are considered as outliers or anomalies. Detecting irregular behavior patterns is sometimes more interesting in some applications~\citep{han2022data}. \citet{karadayi2020unsupervised} conducted anomaly detection in the spatial-temporal COVID-19 data based on the reconstruction error using an autoencoder framework. They first constructed a multivariate spatial temporal data matrix with 20 features including region code, region name, and behavioral attributes such as tests performed, and hospitalized with symptoms. Next, they use a 3D convolutional neural network~\citep{ji20123d} to encode the spatial temporal matrix to preserve both the spatial and temporal dependencies. They then used a convolutional LSTM network~\citep{shi2015convolutional} to decode the previous latent representation. This framework was trained to minimize the reconstruction error where the mean absolute error loss function was employed. After training the network, they calculated the reconstruction error on the test set. A bigger error indicates an anomaly. Experimental results show that the proposed autoencoder framework outperforms other state-of-the-art algorithms in anomaly detection.

\subsubsection{Network Analysis}
Network analysis has shown wide applicability in multiple areas such as biology, social science, and computer networks in analyzing data structures that emphasize complex relations among different entities~\citep{newman2018networks}. It has also been used to analyze human behavior in the context of COVID-19. \citet{shang2021impacts} leveraged a complex network~\citep{barrat2004architecture} which models edges of a network with proportional weights in terms of the intensity or capacity among different elements to investigate bike-sharing behavior under the influence of the pandemic. The data consist of 17,761,557 valid orders and associated travel data of three major dockless bike-sharing operators in Beijing. The study areas were divided into small lattices with the size of $1\:km \times 1\:km$. These lattices are the nodes. If there are bike-sharing trips between two nodes, an edge is drawn. The edge weight is assigned based on the number of bike-sharing trips. They found that the trip distributions in the complex networks are significantly different at various pandemic stages. People are less likely to visit places that were previously popular.

\subsection{Challenges and Opportunities}
Even though most methods we discuss in this section are well-studied, there are several challenges and opportunities when they are applied in COVID-19 studies:
\begin{itemize}
    \item As we have shown in this section, the size of study samples can be up to millions~\citep{shang2021impacts}. Both metric-based (e.g., mutual information, ANOVA) and model-based (e.g., sensitivity analysis) methods may suffer from computational complexity~\citep{christopher2002identification, merz1992measuring}. For instance, many sensitivity analysis frameworks need to iteratively change the input values and re-evaluate the model performance, which is both time-consuming and computationally expensive. One of the important directions for future work can be designing scalable and robust analysis methods. 
    \item Most studies that analyze human behavior use a cross-sectional design that can only examine association but cannot infer causation~\citep{yang2020consumption, ranjit2021covid, vazquez2021impact, lyu2022social}. Causal inference models~\citep{pearl2010causal} that discover the causal effects among variables can be applied more to provide insights into policy design for pandemic control and prevention.
\end{itemize}

\section{Using Big Data to Leverage Human Behavior in COVID-19}
The section intends to answer the research question that whether or not human behavior can be \textit{leveraged} to improve studies that 
address COVID-19 related problems. In particular, we focus on one fundamental area - epidemiology which investigates ``the distribution and determinants of disease frequency''~\citep{macmahon1970epidemiology}. Evidence has shown that NPIs and the development of the pandemic can lead to human behavioral changes~\citep{yan2021measuring, bayham2015measured}, suggesting the desire for mathematical models that can integrate human behavioral factors to better characterize the causes and patterns of the spread of the disease. We primarily discuss three types of models including compartmental models, machine learning models, and hybrid models.

\subsection{Compartmental Model}
Compartmental models divide the population into compartments and express the transition rates between different compartments as derivatives with respect to the time of compartment sizes. SIR is a compartmental model where the population is divided into susceptible class, infective class, and removed class~\citep{brauer2008compartmental}. There have been many efforts of integrating human behavior and other socio-economic factors into the compartmental models to forecast the growth of COVID-19 cases~\citep{chen2020introduction,iwata2020simulation, goel2020mobility, silva2022complex, usherwood2021model}. \citet{chen2020introduction} focused on analyzing the infection data of Wuhan, China. The parameters of the compartment model they chose are from \citet{tang2020updated}. To evaluate the performance of their model, they used real data from Wuhan for simulation. Instead of using parameters of other studies, \citet{iwata2020simulation} designed and enumerated 45 scenarios in terms of the different combinations of their hypothesized parameters, and simulated each.

Motivated by the characteristic that COVID-19 can be spread by both symptomatic and asymptomatic individuals, \citet{chen2020introduction} adopted the SEIR (Susceptible–Exposed–Infectious–Recovered) model by incorporating the migration of the asymptomatic infected population. The proposed SEIAR (Susceptible–Exposed–Infectious–Asymptomatic–Recovered) model was used to analyze and compare the effects of different prevention policies. They found that lowering the migration-in rate can decrease the total number of infected people.

\subsection{Machine Learning Model}
Machine learning models including conventional models such as decision tree, support vector machine, and deep learning models such as long short-term memory network (LSTM)~\citep{hochreiter1997long} intend to learn a function that can recognize patterns or make predictions from data automatically~\citep{han2022data}. Researchers have been employing machine learning models to integrate human behavior to forecast confirmed cases due to its ease of inputting multiple features~\citep{rabiolo2021forecasting, tripathy2021prediction, ramchandani2020deepcovidnet, rashed2021one}. \citet{rabiolo2021forecasting} developed a feed-forward neural network autoregression that integrates Google Trends of searches of symptoms and conventional COVID-19 metrics to forecast the development of the COVID-19 pandemic. The COVID-19 data from the COVID-19 Data Repository~\citep{dong2020interactive} are composed of daily new confirmed cases, the cumulative number of cases, and the number of deaths per million for all available countries. The Google Trends API was used to crawl the search keywords for the most common COVID-19 signs and symptoms. Twenty topics were detected based on the most frequent signs and symptoms of COVID-19. The model that integrates the search data outperforms the model that does not include Google searches. \citet{rashed2021one} designed a multi-path LSTM network that takes mobility, maximum temperature, average humidity, and the number of recorded COVID-19 cases as the input to predict the growth of COVID-19 cases.

\subsection{Hybrid Model}
To enhance the performance of the predictive model, researchers have been proposing to design hybrid models to exploit the compartmental model's ability to model the epidemiological characteristics and the machine learning model's ability to handle various behavioral features~\citep{alanazi2020measuring, bhouri2021covid, aragao2021national}. \citet{bhouri2021covid} used a deep learning model to obtain a more accurate estimation of the parameters of the compartmental model (i.e., SEIR). The susceptible to infectious transition rate was modeled by an LSTM network as a function of Google's mobility data~\citep{google2020covid} and Unacast's social behavior data~\citep{unacast2020unacast} for all 203 U.S. counties. To infer the learnable parameters of the LSTM model, they formulated a multi-step loss function which is expressed as the difference between the variable estimated only from available data and the variable estimated from the learnable parameters. The enhanced SEIR model achieved accurate forecasts of case growth within a period of 15 days. In contrast to \citet{bhouri2021covid} that used deep learning models to improve compartmental models, \citet{aragao2021national} employed compartmental models to aid deep learning models. More specifically, \citet{aragao2021national} used the SEIRD (Susceptible-Exposed-Infected-Recovered-Dead) model to calculate the reproduction rate which was further fed into an LSTM model along with other behavioral indicators to predict COVID-19 deaths. The model using the number of cases and the reproduction number estimated by the compartmental model outperforms the model using only the number of cases by a margin of 18.99\% in average RMSE (Root Mean Squared Error), suggesting the effectiveness of the aid of the compartmental models.  

\subsection{Challenges and Opportunities}
After reviewing the big data methods that leverage human behavior for epidemiology, we summarize several challenges and potential opportunities:
\begin{itemize}
    \item Conventional compartmental models have merits, however, their feasibility is limited due to the strict assumptions for the models. This issue is more critical for the robustness of their applications during the early periods of the COVID-19 pandemic since many parameters regarding the models were unknown. Moreover, real-world situations might be more complicated. For instance, many of these models assumed there would be no nosocomial outbreak of COVID-19 assuming adequate control methods have been conducted. However, nosocomial outbreaks can even happen in developed nations~\citep{varia2003investigation, parra2014first}. The dependence of the conventional compartmental models on strict assumptions may reduce their adaptability to new scenarios (e.g., new variants).
    \item More efforts in considering the mutations of COVID-19 are required because of the high transmissibility~\citep{starr2021prospective}. More importantly, different levels of the perceived risks with respect to various mutations may lead to behavioral changes. 
    \item Different from the conventional compartmental models, machine learning models are data-driven. In other words, the performance and robustness are highly dependent on the amount of available data. For instance, \citet{bhouri2021covid} showed that the models trained for the counties with more COVID-19 cases perform relatively better in estimating the reproduction number. Although it is more desirable since the spread of the disease in regions with more cases may evolve more rapidly, it is also critical to ensure the model performance in regions with fewer reported cases which might be a result of asymptomatic carriers of the COVID-19 virus~\citep{gaskin2021geographic}. There are multiple strategies that can be applied to alleviate the small data issue such as few-shot learning~\citep{wang2020generalizing}. Few-shot learning aims to generalize the model with prior knowledge to new tasks containing only a small amount of data~\citep{wang2020generalizing}. This technique has wide applicability in the COVID-19 context. For instance, when a new variant surfaces in a region, models trained with a large amount of data in other regions can be adopted and generalized to this new scenario which provides the opportunity to model the development of the spread in a more timely manner.
    \item The integration of compartmental models and machine learning models is rather shallow, which is also observed by \citet{xu2021artificial}. Either the compartmental models use the parameters estimated by machine learning models, or the machine learning models take the values predicted by the compartmental models as input data. More sophisticated integration should be investigated. Ensemble methods can be employed to enhance the model performance. Moreover, deep learning models that mimic compartmental models are worth investigating.
\end{itemize}

\section{Discussions}
There are several challenges and open questions that are important to all three mechanisms we have discussed. Understanding and addressing these problems can improve the big data methods for learning human behavior during the time of COVID-19.

\subsection{Data Bias}
Data bias in big data studies on learning human behavior mainly comes from two sources: (1) the selection of the study group, and (2) the method of data collection. For instance, questionnaire-based studies often use convenience sampling to recruit participants where the majority might share some common characteristics (e.g., occupation, geographic location). Social media-based studies can reach more study samples. However, there are demographic differences in the users of different social platforms. For instance, Twitter is found to have fewer older adults.\footnote{\url{https://www.statista.com/statistics/283119/age-distribution-of-global-twitter-users/}} In terms of data collection, questionnaires that collect people's response actively may introduce social desirability bias~\citep{krumpal2013determinants}.

\subsection{Data Privacy}

Big data has been crucial in building applications to fight the pandemic, yet at the expense of privacy~\citep{dataprivacysetting}. With the huge inflow of data via contact tracing, social media, surveillance applications, and so on; there is an increased privacy concern. \citet{aygun2021aspect} trained a BERT (Bidirectional Encoder Representations from Transformers) based model to identify vaccine sentiment on a public dataset released by \citet{bandadataset}. The methodology can be reproduced and used to influence people who are potentially vaccine-hesitant. \citet{cyberattackCTA} stated that Contact Tracing Applications (CTAs) are vulnerable to cyber-attacks and discussed how they could be used for the identification of infected individuals. Authorities that have access might take advantage of this situation and retain data to use in other scenarios that deem fit. This poses a threat to confidentiality and increases the risk of data misuse~\citep{governmentexploit}.

The approach toward data privacy varies across domains and applications. \citet{fahey} discussed two approaches specific to CTAs - data-first that provides autonomy to the authorities to identify individuals in a given jurisdiction, and privacy-first that provides users control of their data and the freedom to share it redacting personally identifiable information. The approaches described here are not binary, and different countries adopt variations of these approaches. 

For social media data, even with the absence of user information associated with their posts, the content acts as a digital fingerprint that can be traced back to the user. \citet{amazonDifferential} provided a methodology to thwart the textual content, preserving its original meaning based on \emph{differential privacy} (DP) in order to tackle this scenario.

\subsection{Interactive Effect}
Most existing studies on learning human behavior focus on the relationship between human behavior and other variables. Future work can pay more attention to the relationship and interactive effects of different individuals. Answering research questions such as whether or not different individuals change behavior due to each other's behavior, and how great the interactive effect is can potentially drive the development of human behavior studies.

\section{Conclusion}
In this paper, we present a systematic overview of the existing studies that investigate and address human behavior within the context of the COVID-19 pandemic. Depending on the mechanism, we categorize them into three groups including using big data techniques to measure, model, and leverage human behavior. We show the wide applicability of big data techniques to studying human behavior during the COVID-19 pandemic. We start with discussions about the studies that measure human behavior in the context of COVID-19. In particular, we observe six major types of data including textual, visual, audio, sensor, interaction, and multimodal. There are several challenges in measuring human behavior because of the three characteristics of the spread of COVID-19 (i.e., global influence, disparate impacts, and rapid evolution). We highlight the influences of these characteristics on the shift in data distribution, model adaptation, fairness, and the feasibility of real-time deployment via techniques such as knowledge distillation. Next, we elaborate on the studies that use big data techniques to model and analyze human behavior. Metric-based (e.g., mutual information score) and model-based methods (e.g., sensitivity analysis) are discussed. We argue that due to the high computational cost of model-based methods, especially when there are hundreds of features, scalable and robust analysis methods are needed. Moreover, we underline the importance of causality analysis in modeling human behavior. We then discuss how big data techniques can be used to leverage human behavior in epidemiology. The models are categorized into compartmental, machine learning, and hybrid. Challenges observed in the studies of these three lines of approaches (e.g., strict assumption, small data) are thoroughly discussed. Finally, we emphasize several challenges and open questions (e.g., data bias, data privacy) that are critical to all three mechanisms we have discussed. Although there are still challenges, we believe that big data approaches are powerful while at the same time, there is room for improvement. The methods and applications surveyed in this study can provide insights into combating the current pandemic and future catastrophes.

\bibliography{report}

\end{document}